\documentstyle[aps,eqsecnum,pre,multicol]{revtex}

\begin{document}

\title{\bf Absence of simulation evidence for critical depletion in slit-pores}

\author{Nigel B. Wilding}
\address{Department of Physics and Astronomy, The University of Edinburgh,\\
Edinburgh EH9 3JZ, U.K.}
\author{Martin Schoen}
\address{Fachbereich Physik--Theoretische Physik, Bergische Universit\"{at} Wuppertal,\\ Gau\ss strasse 20,
D-42097 Wuppertal, Germany}

\input epsf

\maketitle

\begin{abstract} 

Recent Monte Carlo simulation studies of a Lennard-Jones fluid confined
to a mesoscopic slit-pore have reported evidence for ``critical
depletion'' in the pore local number density near the liquid-vapour
critical point. In this note we demonstrate that the observed depletion
effect is in fact a simulation artifact arising from small systematic
errors associated with the use of long range corrections for the
potential truncation. Owing to the large near-critical compressibility,
these errors lead to significant changes in the pore local number
density. We suggest ways of avoiding similar problems in future studies
of confined fluids.

\end{abstract}




\section{Introduction}

In recent papers\cite{SCHOEN95,SCHOEN97,THOMMES95}, one of us has
reported grand canonical Monte Carlo simulation studies of a
Lennard-Jones fluid confined between two structureless attractive walls
arranged in a slit-pore geometry. The behaviour of the number density
profile across the pore, $\rho(z)$, was studied for various values of
the thermodynamic parameters, namely the chemical potential, $\mu$, and
temperature, $T$. At certain values of $\mu$ and $T$ (apparently close
to those of the bulk liquid-vapour critical point) it was observed that
the average local density in the middle of the pore fell markedly below
the value obtained in a fully periodic simulation performed at the {\em
same} $\mu$ and $T$. These findings were used to argue in favour of the
existence of a generic ``critical depletion'' phenomenon, namely the
proposed tendency of a critical fluid to be expelled by a confining
medium, even when the confining walls strongly attract the fluid
particles \cite{SCHOEN97}. Such a scenario is supported by experimental
findings for SF$_6$ adsorbed in mesoporous materials
\cite{THOMMES95,THOMMES94}, for which a dramatic reduction in
adsorption was observed as the bulk critical temperature was approached
from above along the critical isochore.

In this note we point out that the apparent critical depletion
reported in references~\cite{SCHOEN95,SCHOEN97,THOMMES95} is actually a
simulation artifact arising from systematic errors associated with the
corrections applied to the configurational energy to compensate for the
truncation of the interparticle potential. Using new simulations, we
show that if one chooses a sufficiently large truncation distance or 
alternatively avoids the use of truncation corrections altogether,
then the depletion effect disappears.

\section{Simulation details and results}

The simulation arrangement and procedure employed in this work are the
same as those described in refs.~\cite{SCHOEN95,SCHOEN97}, and accordingly
we merely summarise the principal features. 

Grand canonical Monte Carlo simulations \cite{FRENKEL} were performed for a
Lennard-Jones fluid, having an interparticle potential of the form:

\begin{equation}
U_{LJ}(r)=4\epsilon\left[\left (\frac{\sigma}{r}\right)^{12}-\left(\frac{\sigma}{r}\right)^{6}\right ]\;.
\label{eq:lj}
\end{equation}
where $\epsilon$ and $\sigma$ are respectively the Lennard-Jones well depth
and scale parameters. Two distinct geometries were studied:

\begin{enumerate}

\item A fully periodic system.

\item A slit-pore geometry, in which the fluid is confined between two
parallel structureless walls, having periodic boundary conditions in
the directions parallel to the walls.

\end{enumerate}
In the latter case, the walls were taken to exert a potential on the
fluid particles of the form:

\begin{equation}
U_{FW}=4\epsilon f\left[ \frac{2}{5}\left(\frac{\sigma}{z}\right)^{10}-\left(\frac{\sigma}{z}\right)^{4}\right ]\;,
\end{equation}
where $f$ is a parameter that tunes the strength
of the wall-fluid interactions relative to those of the fluid
interparticle interactions.

As in the previous studies of this system \cite{SCHOEN95,SCHOEN97}, the
reduced temperature was set to the value $T=1.36$, believed to be close to the
bulk critical temperature. The chemical potential $\mu$ of the periodic
system was then tuned until the equilibrium density reached the value
$\rho=0.365$, believed to be close to the bulk critical density. The
resulting value of $\mu$ was then fed into a simulation of the
slit-pore system at the same temperature and with the choice
$f=0.9836$. In both the periodic and slit-pore arrangements, the
Lennard-Jones interparticle potential was truncated at some radius and
a compensating correction applied to the configurational energy. For
the periodic system, this correction was calculated in the standard
fashion by assuming a spherical cutoff surface of radius $r_c$ centred
on each particle, combined with a uniform density approximation (UDA) for
$r>r_c$. This yields for the energy correction per particle:

\begin{equation}
u_{pbc}=\frac{1}{2}4\pi\rho\int_{r_c}^\infty dr r^2 U_{LJ}(r)=\frac{8}{3}\pi\rho\epsilon\sigma^3\left[\frac{1}{3}\left (\frac{\sigma}{r_c}\right)^9-\left (\frac{\sigma}{r_c}\right)^3\right ]\;,
\end{equation}
where $\rho=\langle N\rangle/V$ is the average number density of the
system. 

In the case of the slit-pore system, the truncation correction used was
that given by Schoen et al. \cite{SCHOEN87}, which assumes a
cylindrical cutoff surface (on whose principal axis each particle lies)
extending across the whole width $D$ of the pore, i.e. such that the
cylinder ends coincide with the pore walls. This yields

\begin{equation}
u_{pore}=-\frac{\pi\epsilon\sigma^6}{s_c^3}\tan^{-1}\left (\frac{D}{s_c}\right)V\rho^2 \;,
\label{eq:cylcor}
\end{equation}
where $s_c$ is the radius of the cutoff cylinder.

We have studied the effect of the cylindrical cutoff radius $s_c$ on
the density profile $\rho(z)$ of the slit-pore system. For each choice
of $s_c$, the $\mu$ value employed in the simulation was that yielding
an average density $\rho=0.365$ in a periodic system of the same
dimensions and with $r_c=s_c$. For small cylindrical cutoffs
($s_c=3.5\sigma$), figure~\ref{fig:cutdep} shows that the local density
in the pore middle is depleted with respect to the density of the
periodic system (dashed line) at the same $T,\mu$. This is the result
reported in refs.~\cite{SCHOEN95,SCHOEN97,THOMMES95}. However, new
results for a larger choice of the cylindrical cutoff (also included in
figure~\ref{fig:cutdep}) show that this depletion reduces as $s_c$ is
increased, and in fact vanishes for $s_c\gtrsim 5.0\sigma$. This
dependence of the depletion on the choice of $s_c$ was missed in the
previous studies \cite{SCHOEN95,THOMMES95}.

We have also performed simulations in which we dispense with the use of
cutoff corrections altogether and simply simulate a system of particles
interacting via a truncated Lennard-Jones potential. The results
(figure~\ref{fig:notrunc}), exhibit no sign of a density depletion in
the pore middle with respect to the periodic system.

\section{Discussion and conclusions}

\begin{figure}[h]

\setlength{\epsfxsize}{8.0cm}
\centerline{\mbox{\epsffile{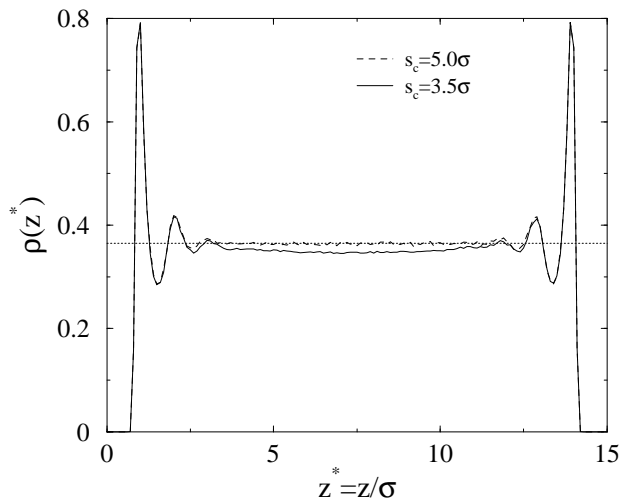}}}

\caption{The number density profile $\rho(z)$ (corresponding to a
slit-pore of width $D=15\sigma$) at two values of the cylindrical
cutoff radius. In each case, the reduced temperature is $T=1.36$ and
the chemical potential $\mu$ used is that which yields an average
density $\rho=0.365$ (horizontal line) in a periodic system of linear
size $L=15\sigma^3$ at the same $T$. The figure shows that as the
cylindrical cutoff radius is increased, the depletion disappears.}

\label{fig:cutdep}
\end{figure}

\begin{figure}[h]
\setlength{\epsfxsize}{8.0cm}
\centerline{\mbox{\epsffile{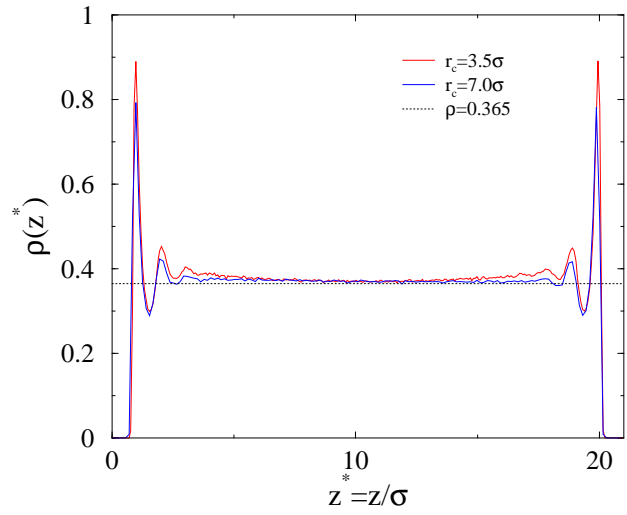}}}

\caption{The number density profile $\rho(z)$ (corresponding to a
slit-pore of width $D=21\sigma$) for a truncated potential with no tail
correction applied. In each case the reduced temperature is $T=1.36$
and the chemical potential $\mu$ used is that which yields an average
density $\rho=0.365$ (horizontal line) in a periodic system of linear
size $L=21\sigma^3$ at the same $T$.} 

\label{fig:notrunc} 
\end{figure}

The dependence of the density depletion on the choice of the
cylindrical cutoff $s_c$ (figure~\ref{fig:cutdep}) points to a
breakdown of the uniform density approximation invoked in the
derivation of the truncation correction for the internal energy,
eq.~\ref{eq:cylcor}. This approximation assumes that the number density
outwith the cutoff surface is uniform, having the average system
density $\rho=\langle N \rangle$/V. However, figures~\ref{fig:cutdep}
and \ref{fig:notrunc} show that for a slit system, $\rho(z)$ exhibits
considerable structure across the pore, especially close to the walls.
Accordingly, one must expect some measure of systematic error to be
associated with eq.~\ref{eq:cylcor}. Tests show that for the choice of
cutoff $s_c=3.5\sigma$ employed in references
\cite{SCHOEN95,SCHOEN97,THOMMES95}, this error is very small, so that
in most circumstances eq.~\ref{eq:cylcor} represents a good
approximation.

It seems, however, that in the critical region, even a very small
error in the truncation correction can lead to large effects on the
local pore number density. The reason for this is the large
near-critical compressibility, reflected in the fact that near $T_c$,
isotherms of $\mu(\rho)$ become very flat (see eg. figure $3$ of
\cite{SCHOEN97}). Since the error in the truncation correction acts
rather like a shift in the bulk (chemical potential) field with respect
to the periodic system, large alterations to the local pore density may
result. This is in accord with the observation
\cite{SCHOEN95,THOMMES95} that the depletion is large close to the
critical point, but diminishes as one moves to higher temperatures
along the critical isochore.  

To avoid similar problems arising in future studies of the effects of
confinement on near critical fluids, it would seem wise to adopt one 
of the following strategies:

\begin{enumerate}

\item Employ a very large value for the truncation range and test for
any systematic dependence of results on its value. This is clearly a
very computationally intensive solution.

\item Employ a truncated potential and dispense with corrections
altogether. Such a system is clearly well defined but causes
complications if one wishes to model real substances.

\item Employ a cut and shifted potential which tends smoothly to zero
at the cutoff. 

\end{enumerate}

In summary, we have demonstrated that the apparent critical depletion
reported in \cite{SCHOEN95,SCHOEN97,THOMMES95} was actually an artifact
arising from systematic errors in the energy correction for the tail
truncation in the slit-pore geometry. Although numerically small, these
errors can (in the critical region) strongly influence the fluid local
number density of the confined system compared to a periodic system at
the same temperature and chemical potential. Thus our findings
underline the care that must be taken when implementing any sort of
truncation corrections for near critical fluid models.

Further simulation studies of the near-critical properties of a
confined fluid are in progress and a detailed account of these and
their implications for theory and experiment on critical depletion will
be presented in a later paper.

\acknowledgements

NBW is grateful to R. Evans and A. Maciolek for stimulating his
interest in critical depletion, and for most useful correspondence. He
also thanks the Royal Society (grant number 19076), the Royal Society
of Edinburgh and the EPSRC (grant no. GR/L91412) for financial support.
MS is grateful to G.H. Findenegg for discussions and acknowledges
financial support from the Sonderforschungsbereich 448 ``Mesoskopisch
strukurierte Verbundsysteme''

\end{document}